\documentclass[11pt]{article}                  
  \usepackage{times}                   

 \usepackage[dvips]{graphicx}

\begin{document}
\title{Multi-scale magnetic field intermittence in the plasma sheet}

\author{Z. V\"{o}r\"{o}s (1), 
W. Baumjohann (1),
R. Nakamura (1),A. Runov (1),\\ T. L. Zhang (1), M. Volwerk (1), H. U. Eichelberger (1), \\
A. Balogh (2), T. S. Horbury (2), \\
K.-H. Gla\ss meier (3), \\
B. Klecker (4) \\
and H. R\`{e}me (5) \\
(1) Institut f\"{u}r Weltraumforschung der \"{O}AW, Graz, Austria,\\
(2) Imperial College, London, UK, \\
(3) TU Braunschweig, Germany, \\
(4) Max-Planck-Institut f\"{u}r extraterrestrische Physik, Garching, Germany,\\
(5) CESR/CNRS, Toulouse, France. 
}
\date{}
\maketitle

\begin{abstract}
This paper  demonstrates that intermittent magnetic field  fluctuations 
in the plasma sheet exhibit transitory, localized, and multi-scale features. We propose
a multifractal based algorithm, which quantifies intermittence on the basis of the 
statistical distribution of the
'strength of burstiness', estimated within a sliding window. Interesting multi-scale phenomena observed
by the Cluster spacecraft include large scale motion of the current sheet and bursty bulk flow associated
turbulence, interpreted as a cross-scale coupling (CSC) process.
\end{abstract}

\section{Introduction}\label{sec:intro}

The study of turbulence in near-Earth cosmic plasma is important in many respects. Turbulence, 
being in its nature a multi-scale phenomenon, may influence the transfer processes of energy, mass
and momentum on both MHD and kinetic scales. Vice versa, turbulence can be driven by 
instabilities such as  magnetic reconnection or current disruption  
\cite{tetr92, ang99a, klim00, chan02, lui02}  

The understanding of intermittence features of fluctuations is fundamental to  turbulence.
Intermittence simply refers to processes which display 'sporadic activity' during only a small 
fraction of the considered time or space.  This is also the case in non-homogeneous turbulence 
where the distribution of energy dissipation regions is sporadic and probability distributions
of measurable quantities are long-tailed with significant departures from gaussianity. Rare events
forming the tails of probability distribution functions, however, carry a decisive amount of energy
present in a process \cite{fris95}.

Substantial experimental evidence exists for the occurence of intermittent processes within the plasma sheet.
\cite{baum90} showed that within the inner plasma sheet inside of 20 $R_E$ high-speed
short-lived ($\sim$ 10 sec) plasma flows are rather bursty. \cite{ange92} noted
that those flows  organize themselves into $\sim$ 10 min time scale groups called bursty bulk flows (BBF).
Despite the fact that BBFs represent relatively rare events (10-20 \% of all measurements), they are the 
carriers of the decisive amount of mass, momentum and magnetic flux \cite{ang99b, scho01}
and 
can therefore energetically influence the near-Earth auroral regions \cite{naka01}.

So far experimental evidence for real plasma sheet
turbulence is not unambiguous, however its existence is supported by
the occurrence of plasma  fluctuations  in  bulk
flow velocity and magnetic field which are comparable or even larger than the corresponding mean 
values \cite{boro97}. Other characteristics of plasma sheet turbulence, such as probability
distributions, mixing length, eddy viscosity, power spectra, magnetic Reynolds number, etc., were found
to exhibit the expected features or to be in expected ranges predicted by turbulence theories 
\cite{boro97}.
Though the amplitude of the velocity and magnetic field fluctuations 
increases with geomagnetic activity \cite{neag02}, intense fluctuations are present independently
from the level of geomagnetic activity \cite{boro97},  indicating that 
different sources or driving mechanisms
might be involved in their generation. In fact, according to observations by \cite{ang99a},
at least a bi-modal state of the inner plasma sheet convection is recognizable from plasma flow magnitude
probability density functions: BBF--associated intermittent jet turbulence and intermittent turbulence 
which occurs during non-BBF (quiet background) flows. \cite{ang99a} have 
also proposed that BBF--generated intermittent turbulence can alter transport processes in plasma
sheet and may represent a way that cross-scale coupling (CSC) takes place.

These facts call for a method which  allows  analysis of both intermittence and multi-scale properties
of fluctuations.
In this paper we propose  a multifractal technique for this purpose.
Using both magnetic field and ion velocity data from Cluster, we will show  that
BBF--associated 'magnetic turbulence'  exhibits clear signatures of cross-scale energisation.

\section{Multifractal approach to turbulence}
\label{sec:mul}

In order to elucidate  the basic assumptions of our approach we use a multinomial distribution 
model first and introduce a local parameter for quantification of the intermittence level on a given scale.
Then we discuss the  range of potential scales over which the presence of cross-scale energisation 
might be experimentally demonstrable and mention  some limitations regarding the availability of 
multipoint observations.

\subsection{Local intermittence measure ($LIM$)} 
The large scale representation of magnetotail processes by mean values of measurable quantities is useful
but can also be misleading in characterising  multi-scale phenomena when quantities observed 
on different scales carry physically important information.

Multifractals are well suited for describing  local scaling properties of dissipation fields in 
non-homogeneous turbulence \cite{fris95}. Therefore they are most suitable 
for a description of plasma sheet fluctuations. 
In non-homogeneous turbulence, the transfer of energy
from large scales to smaller scales can be conveniently modeled by a multiplicative cascade process. 
The distribution of energy dissipation fields on small scales exhibits burstiness
and intermittence. 

Let us consider a simple model example. Multinomial deterministic measures are examples of 
multifractals \cite{ried99}.
These consist of a simple recursive construction rule: a uniform measure $\mu (L)$ is chosen on an interval
$I:[0,L]$ and is then unevenly distributed over $n > 1$ ($n$ - integer) equal subintervals of $I$
using weights $m_i$; $i=1,...,n$ and $\sum_{i}m_{i}=1$. Usually $L$ is chosen to be 1. 
After the first iteration we have $n$ equal 
subintervals, and  subinterval $i$ contains a fraction $\mu (L) m_i$  of $\mu (L)$. Next every 
subinterval and the measure on it are splitted in  the same way recursively, having $i=1,...,n^k$
subintervals or boxes after $k$ iteration steps and  $\mu_{k,i}$ in the box $I_{k,i}$. Figure 1
shows the simplest example of a binomial distribution ($n=2$). We note that the measure $\mu $ can be any 
positive and additive quantity, such as energy, mass, etc.

Figure 2a presents two distributions, $A$ and $B$, separated by a dashed vertical line in the middle.
Both mimic typical bursty 'time series' like a physical variable from a turbulent system, however, 
by construction
distribution $A$ is less intermittent than  distribution $B$ . In both cases
the same initial mass ($\mu$) is distributed over interval $L$, 
$n=8$; $k=5$ is chosen (that is $n^{k}=32768$ boxes), but
the weights $m_{i}(A) = (0.125, 0.08, 0.09, 0.16, 0.05, 0.25, 0.12, 0.125)$ and
$m_{i}(B) = (0.1, 0.3, 0.05, 0.002, 0.04, 0.218, 0.09, 0.2)$ are different.
Intermittence is larger in case $B$ 
(Figure 2a) because of the larger differences between weights (if all weights were
equal, the resulting distribution would become homogeneous). Our goal is to quantify this level of 
intermittence by multifractals. The definition of multifractality in terms of the large deviation principle 
simply states that
a dissipation field, characterized locally by a given 'strength of burstiness' $\alpha$, has a distribution
$f(\alpha)$ over the considered field. It measures a deviaton of 
the observed  $\alpha$ from the expected value ${\overline \alpha }$. 
The corresponding ($\alpha, f(\alpha $)) large deviation spectrum is of concave
shape \cite{ried99}.

The strength of local burstiness, the so called coarse-grain H\"{o}lder exponent $\alpha$, is
computed as
\begin{equation}
\alpha_i \sim \frac{log \mu _{k,i}}{log [I]_{k,i}}
\end{equation}
where $[I]_{k,i}$ is the size of the  $k,i$-th box and equality holds asymptotically.

It is expected that due to its multiplicative construction rule $\mu_{k,i}$ will decay fast as 
$[I]_{k,i}\rightarrow 0$ and $k\rightarrow \infty$. We add that $\alpha _{i}<1$ indicates bursts on all scales 
while $\alpha _{i}>1$ characterizes regions where events occur sparsely \cite{ried99}.
Equation (1) then expresses the power-law dependence of the measure on resolution. Usually 'histogram methods'
are used
for the estimation of the $f(\alpha )$ specturm (called also rate function), 
so that the number of intervals $I_{k,i}$ for which
$\alpha_{k,i}$ falls in a box between $\alpha_{min}$ and $\alpha_{max}$ (the estimated minimum and maximum 
values of $\alpha$) 
is computed and $f(\alpha )$ is found
by  regression. In this paper, however, $f(\alpha )$ spectra are estimated using the FRACLAB package
which was developed at the Institute National de Recherche en Informatique, Le Chesnay, France. Here the well 
known statistical kernel method for density estimations  is used which also yields satisfactory estimations
for processes different from purely multiplicative ones \cite{vehe98, canu98}.

A comparison of Figures 2a and 3 indicates that
the wider the $f(\alpha )$ spectrum the more intermittent  the measure. This  feature  was also 
proposed to study the possible role of turbulence in solar wind - magnetosphere coupling processes
\cite{voro02} and this feature will be used to describe magnetic field intermittence in the plasma sheet.

In order to gain appropriate information about the time evolution of intermittence from real data we estimate
$f(\alpha )$ within sliding overlapping windows $W$ with a shift $S\ll W$. In our model case the time axis is
represented by increasing number of subintervals $I_{k,i}$. $LIM$ is introduced as the total area under each
$f(\alpha )$ curve within a window $W$,  divided by the mean area obtained from the measurements along the 
reference measure $A$. 
Actually $LIM(A)$ 
fluctuates around 1 
due to  errors introduced by finite window length. For measures, exhibiting higher level of intermittence than
the reference measure $A$, $LIM> 1$. 
Figure 2b shows that for measures $A$ and $B$ the different levels of intermittence are 
properly recognized by $LIM$. Estimations based on a larger window (Window I: $W=7000$ boxes, $S=100$ boxes)
are more robust, but a smaller window (Window II:$W=2000$ boxes, $S=100$ boxes) allows a better localization
of the transition point between measures $A$ and $B$ (thick line in the middle of Figure 2a).

\subsection{Multi-scale $LIM$}

Deterministic multinomial measures are self-similar in the sense that the construction rule is the same at 
every scale. Real data are more complex. Physical processes may have characteristic scales or may distribute 
energy differently over some ranges of scales. In order to study BBF--associated magnetic 
turbulence on both large
and small scales we introduce a 'time scale' $\tau$ through differentiation
\begin{equation}
\delta B_{x}(t,\tau ) = B_{x}(t+\tau ) - B_{x}(t)
\end{equation}
Throughout the paper the GSM coordinate system is used in which the x-axis is defined along the line
connecting the center of the Sun to the center of the Earth. The origin is defined at the center of
the Earth and is positive towards the Sun.
Then a normalized measure at a time $t_i$ is given by
\begin{equation}
\mu_{B_x}(t_{i},\tau ) = \frac{\delta B_{x}^{2}(t_{i},\tau )}{\sum_{i}\delta B_{x}^{2}(t_{i},\tau )}
\end{equation}
We have to mention, however, some essential limitations of this approach when  a separation
of spatial and temporal variations is eventually addressed. A time series obtained from a single spacecraft 
can be used for mapping the spatial structure of turbulence using the so called Taylor's hypothesis if
the spatial fluctuations on a scale $l$ pass over the spacecraft faster than they typically fluctuate in time.
In the plasma sheet this can probably be  the case during fast BBFs \cite{horb00}. 
Otherwise Taylor's hypothesis may not be completely valid. Instead of Equation (2) a real two-point
expression, $\delta B_{x+l}(t) = B_{x+l}(t) - B_{x}(t)$ could be used, where $l$ is a distance between Cluster
spacecraft. The corresponding $LIM$, however, strongly fluctuates in a variety of cases (not shown), presumably
due to mapping of physically different and structured regions by individual Cluster satellites. We postpone
this kind of multi-point observations to future work.

Nevertheless, \cite{ang99a} noticed that some characteristics of turbulence estimated from
single point measurements are equivalent to ones from two-point measurements for distances at or beyond 
the upper  limit of the inertial range in which case Equation (2) can be used efficiently. \cite{boro97}
estimated the lower limit of inertial range to be about ion gyroperiod time scales ($\sim$ 10 sec in
plasma sheet), over which a strong dissipation of MHD structures is expected. The upper limit of inertial 
range (largest scale) was identified by plasma sheet convection time scale or by inter-substorm time scale,
both of order 5 h. As  known, inertial range refers to a range of wavenumbers (or corresponding scales)
over which turbulence dynamics is characterized by zero forcing and dissipation \cite{fris95}.
Recent theoretical and experimental work shows, however, that inertial range cascades might be exceptional. 
In a large 
variety of turbulent flows rather bidirectional direct coupling (or cross scale coupling - CSC) due to nonlinearity and nonlocality 
between large and small scales exists \cite{tsin01}. While the large scales
are determined by velocity fluctuations, the small scales are represented by the field of velocity derivatives
(vorticity, strain).

\section{Data analysis}
\label{sec:dat}

\subsection{General considerations}
In this paper we analyse intermittence properties of 22 Hz resolution magnetic field data from the Cluster (CL)
fluxgate magnetometer (FGM) \cite{balo01} and compare those characteristics with the
spin-resolution ($\sim$ 4 sec) velocity data from the Cluster ion spectrometry (CIS/CODIF) experiment
\cite{reme01}. 

Compared with the previous model example, the estimation of the $LIM$ for the $B_X$  component of the magnetic 
data was 
somewhat different. First of all, we calculated $LIM(t,\tau)$ for different time scales $\tau$. 
In optimal case energization through a cascading process should appear on different scales time shifted,
that is the large scales should become energized first and the small scales later.
We found, however,
that on various scales $LIM$  fluctuates strongly (not shown) and using this 
approach it would be  hard  to identify an
energy cascading process within an inertial range of scales. This was not unexpected, because  
cascade models are treated in Fourier space (wave vector space) whereas our approach represents a pure
time-domain analysis  method (though the magnetic field data itself already contain some spatial information),
so the individual scales have rather different meanings. Also, nonlinear
and nonlocal direct interactions between scales may prevent experimental recognition of cascades.

Therefore, we decided to
estimate $LIM$ on several scales around 40 sec, which is considered to be  a typical large scale of BBF
velocity fluctuations, and compute the average $LIM_L$ (subscript $L$ reads as large scale)
from the corresponding $f(\alpha)$ spectra.
BBF events usually last several minutes \cite{ange92}, however, if $\tau$ is chosen to be 
several minutes long, 
the corresponding window length W should be even several times longer  which would make 
measurements of the non-stationary
features of intermittence almost impossible.

A typical small
scale was chosen experimentally. We  looked for a $\tau$ (Equation 2) which reflects the small scale
changes of the intermittence level properly. We  found that fluctuations on time scales larger than a few seconds
already exhibit similar intermittence properties as on scales around 40 sec. In fact, the majority of bursty 
flows may remain uninteruptedly at high speed levels for a few seconds \cite{baum90}. 
Therefore we considered  time scales around 0.4 sec  as small ones (two orders less than the chosen 
large scale) and the corresponding intermittence measure reads as $LIM_S$. This timescale may already 
comprise some kinetic effects.
The use of 
22 Hz resolution magnetic 
data from FGM experiment on such small time scales implies the problem of different transfer functions for
high and low frequencies. Corrections  introduced by appropriate filtering had no effect on the 
$LIM$ estimations.

\subsection{Event overview and $LIM$ analysis}
The events, we are interested in, occured between 1055 and 1135 UT on August 29, 2001 (Figure 4a), 
when CL
was located at a radial distance of about 19.2 $R_E$, near midnight. 
In the following   the relatively 'quiet' time period from 
1115 to 1120 UT will be used as a reference level for both $LIM_L$ and $LIM_S$  estimations. 
It means, that during this time period the $LIM_{L,S}$ mean values equal 1.

The current sheet structure and movement during 1055 - 1107 UT has been studied by 
\cite{runo02}. 
Only the $B_X$ component  from CL 3 will be evaluated. 
During the chosen interval CL 3
was located approximately 1500 km south of the other three spacecraft. 
CL  traversed the neutral sheet 
from the northern ($B_{X}\sim 20 $ nT) to the southern hemisphere ($B_{X}\sim -15 $ nT), then $B_X$  
approached $B_{X} \sim 0$ again (Figure 4a). The correspondingly normalized small scale ($\tau$=0.4 sec)
and large scale ($\tau$=40 sec) measures (Equations 2 and 3) are depicted by red and blue curves 
in Figures 4 b an c, respectively. In fact, Equaton 2 represents a high-pass or low-pass filter for properly
chosen time shifts $\tau$. Therefore Figure 4b (4c) shows an enhanced level of small-scale (large-scale)
fluctuations when high-frequency (low-frequency) fluctuations are present  in Figure 4a. 
$LIM_{L,S}$ were computed as a changing area under $f(\alpha )$ multifractal distribution curves
over the interval $\alpha \in (1,\alpha _{max})$ and within sliding window W=318 sec.
The time shift is S=4.5 sec.
These parameters were chosen such that the opposing requirements for stability of $LIM$ estimations (wide window
needed) and for time-localization of non-stationary events (narrow window needed) were matched. 
Considering the whole area under the $f(\alpha)$ curves, i.e. estimating 
$LIM$ over  $\alpha \in (\alpha _{min},\alpha _{max})$ as in the  previous section (model case) would be also 
possible. This gives, however, the same qualitative results. During 
intervals of changing intermittence level mainly the right wing of $f(\alpha) $  changes.  
Therefore
we estimated $LIM$ over the interval $\alpha \in (1,\alpha _{max})$.
Figure 4d shows, 10 red curves of $LIM_S(t,\tau )$ 
computed for $\tau \in (0.3, 0.5)$ sec, and 10 blue curves for $\tau \in (30, 50)$ sec. Obviously,
$LIM_L$ and $LIM_S$ exhibit quite different courses and we will analyse the differences in more detail.

First, we examine the $f(\alpha)$ multifractal spectra. Windows A, B, C and D in Figure 4a 
indicate periods during which distinct physical phenomena occured. The differences are evident 
from the magnetic field $B_X$, measures $\mu_{B_x}$ and $LIM_{L,S}$ evolution over time (Figure 4 a--d). 
We focus mainly on an interval between
1123 UT and 1133 UT in which both $LIM_{L}$ and $LIM_{S}$ have increased values. Period C is during this interval.
We contrast this  interval with  1055 to 1110 UT,
at the beginning of which a wavy flapping motion or  an expansion-contraction of the current sheet 
is observed (Period A) with a characteristic time scale of 70-90 sec \cite{runo02}. 
Periods B and C represent quiet intervals with  different $B_X$ values. The corresponding $f(\alpha) $ 
spectra are depicted by red and blue circles in Figure 5.
We also computed the global $f(\alpha)$ spectra for the whole $B_X$ time series 
on small and large scales  from 1055 to 1135 UT, which are depicted
by solid red and blue lines, respectively. Deviations from these average
$f(\alpha)$ curves classify physical processes occurring during periods A--D. An examination of
only the right wings of the distributions leads to the following conlusions (see also Figure 4a and d):
(1.) the $f(\alpha) $ spectra estimated on both large and small scales exceed the average $f(\alpha) $
only during period C; (2.) during period A (large scale flapping motion) only the large scale (blue circles)
exceed the average blue curve significantly; (3.) quiet periods B and D exhibit average or narrower than average 
distributions.

With the definition of $LIM$, we have introduced a number
which quantifies intermittence as an area under the right wing of the $f(\alpha) $ distribution function.
We have to emphasize, however, that $f(\alpha)$ distributions cannot be described or replaced by one number.
The
whole distribution contains more information. It is evident from Figure 5 that the more intermittent period
C is also characterized by the largest difference between $\alpha_{max}$ and $\alpha_{min}$ on
small scale (red circles). Also only in this case the maximum of the $f(\alpha )$ curve is significantly
shifted
to the right. There are multiplicative cascade models for which multifractal distributions of concave
shape and the underlying
intermittence properties
can be described by one parameter, e.g. the P-model \cite{hals86, voro02}. 
However, those models cannot fit the data well because of the non-stationarity
and shortness of  the available time series in the plasma sheet. This is clearly visible in the case of  
large scale non-concave distributions   during periods A and C (blue circles, Figure 5). For this reason
$LIM$ represents a descriptor which tells more about the intermittent fluctuations than second order statistics,
but less than the whole multifractal distribution function.

\subsection{Multi-spacecraft comparison and BBF occurrence}
To facilitate interpretation, the $B_X$ components from two Cluster spacecraft (CL1 and CL3) are depicted
in Figure 6a.
The difference between the $B_{X}$ components measured at the locations
of CL1 and 3 changes substantially during the considered interval, indicating spatial gradients  
of the order of the distance between CLs within current sheet.
The largest spatial gradients occur during and after the flapping motion from 1055 to 1110 UT. Large 
gradients are also present during interval  1122 -- 1130 UT.
These two intervals are separated by a $\sim$ 10 min interval, from 1110 to 1121 UT, characterized by small
spatial gradients and $-18 < B_{X} < -10$ nT. 
Therefore, the spacecraft are 
outside of the current sheet. 
There are two more  periods when the observed spatial gradients are small. The first is 
before 1055 UT ($B_{X} > 18$ nT), 
when the spacecraft were in the northern lobe. The interval after 1130 UT 
contains also small spatial gradients, 
but the $B_{X}$ components change from -6 to 2 nT, indicating that the spacecraft are closer to the center of current sheet.

Figure 6b shows
$LIM_{L,S}$ (red and blue curves).
Standard deviations computed from a number of $f(\alpha )$ distributions (Figure 4d) estimated 
around $\tau = 40$  and $0.4$ sec are also depicted by thin lines round $LIM_{L,S}(t)$ in Figure 6b. 
Window parameters are also indicated.

It is visible that during the  large scale motion (thoroughly analysed by \\
\cite{runo02}) 
and after, until $\sim$ 1110 UT 
(Figure 6a),  $LIM$ shows enhanced intermittence level on large scales, but not 
on small scales (Figure 6b).
$LIM_{L}$ is also high before 1055 UT, only because the  local window W extends over the
period of wavy motion of current sheet. As no enhanced intermittence level is observed during the whole
interval until $\sim$ 1110 UT
on small scales, we conclude that cross-scale energisation is not present. More precisely, at least in
terms of intermittent fluctuations quantified by $LIM$, 
there was no  CSC mechanism present that could couple large scale energy reservoirs at
the level of the MHD flow ($\sim$ 40 sec) to the small  scales ($\sim$ 0.4 sec).
We cannot exclude, however,
other mechanisms of CSC not directly associated with $LIM$ changes.

$LIM_{L}$ tends to decrease rapidly after 1110 UT because data 
from outside the current sheet influence its estimation.

Between 1120 and 1135 UT both $LIM_L$ and $LIM_S$ increase. This enhancement is clearly associated 
with high frequency intermittent fluctuations in $B_X$ (Figure 6a; see also the global spectrum for period C
in Figure 5) and with occurence of a BBF. In Figure 6c we show the  
proton velocity data from CIS/CODIF experiment ($H^{+}V_{X}$; GSM).
Figure 7a shows magnetic field $B_Z$ component of the magnetic field measured
by CL3 while Figure 7b -- d show $B_X$, proton velocity 
and $LIM$ at better time resolution than in Figure 6.

Four windows centered on points 
marked by crosses indicate the times when $LIM_{L,S}$ significantly 
increase or decrease relative to the quiet level
($LIM_{L,S}\sim$1). Vertical red and blue arrows indicate the starting points of increase and decrease 
of $LIM_{L,S}$, respectively.

When the spacecraft enter the current sheet  after 1120 UT, $LIM_L$  increases and window 1 shows that the 
enhancement 
is associated with the appearance of large scale fluctuations in $B_X$, a small decrease of $B_Z$ 
and gradual increase of $V_{X,H^{+}}$ starting at 
1122:20 UT (see the vertical dashed line at the right end of window 1). Approximately two minutes
later, the center of window 2 points at first significant enhancement of $LIM_S$ (red vertical arrow). 
$LIM_S$ achieved its maximum value 1.14 $\pm$ 0.02 within $\sim$ 40 sec.
The right end  of window 2 is clearly associated with: (1.) magnetic field dipolarization (rapid increase of 
$B_Z$ to $\sim$ 8-10 nT in Figure 7a.); (2.) appearance of high frequency fluctuations in $B_{X}$(CL3), 
(in Figure 7b.);
(3.) BBF velocities larger than 400 $km/s$ (Figure 7c.); (4.) enhancements of energetic ion and electron
fluxes on CL3 (not shown); all at $\sim $ 1124:27 UT.  

$LIM_S$ drops to 1.05 $\pm$ 0.02 at 1127:45 UT 
(marked by red arrow from the center of window 3). This time, the right end of window 3 starts to leave behind
the largest peaks of $V_{X,H^{+}}$, but that is not the only reason of the decrease of $LIM_S$. 
When $LIM_S$ decreases, $LIM_L$ remains at high level (1.24 $\pm$ 0.05), or even increases, 
because of the sudden jump in $B_X$ form -10 to +2 nT closely before 1130 UT. It was previously 
mentioned that after 1130 UT the spacecraft got closer to the center of current sheet. Therefore, we 
suppose that due to the large scale motion of the current sheet, which  keeps $LIM_L$ at a high level, 
the spacecraft appear
to be outside of the region of BBF--associated turbulence. This is also supported by the simultaneous decrease
of both $LIM_L$ and $LIM_S$ at approximately 1132:30 UT, when  window 4 includes $B_X$ 
from the region with small gradients after 1130 UT.
Therefore, during the interval between the right ends of window 2 and  3, i.e. within a time period 
of $\sim$ 6 minutes from ~1124 to ~1130 UT,  $LIM$ analysis indicates BBF and dipolarization associated
CSC between MHD and small, possibly kinetic scales. 
An alternative to the CSC might be a simultaneous, but independent 
enhancement of intermittent fluctuations on both large and small scales. As was mentioned earlier, 
an identification of the energy-cascading process is almost impossible using the applied method. 
The primary pile-up of energy associated with increase of BBF velocity on large scales at 1122:20 UT,
however, seems to indicate that in this case small scale fluctuations are energised by MHD scale 
rapid flows. Unambiguous evidence for or against BBF--related CSC requires a statistical ensemble of events
to be analysed. We mention, inverse
cascades during current disruption events were reported by \cite{lui98}.

The large difference between $LIM_L$ and $LIM_S$ 
after 1128 UT can be attributed to the prevailing large scale motion of the current sheet.
The spacecraft got closer to the centre of current sheet where the multiscale $LIM$ signs of CSC
are already absent. This can be explained by  the transitory and localized nature of CSC.

\section{Conclusions}
\label{sec:con}

We proposed a windowed multifractal method to quantify local intermittence of magnetic field fluctuations
obtained by Cluster. The main results of this paper comprise a multi-scale description of
large scale current sheet motion and of a BBF--associated cross-scale energisation process. 
We have shown as Cluster passes through
different plasma regions, physical processes exhibit non-stationary intermittence properties on MHD
and small, possibly kinetic scales. As any robust estimation of turbulence characteristics requires processing of 
long time series (due to the presence of energetic but rare events), the observed 
transitory and non-stationary nature
of fluctuations prevents us to unambiguously support or  reject  a model for plasma sheet turbulence. 

The multifractal description of intermittent magnetic fluctuations is in accordance with previous
knowledge that the change of fractal scaling  properties can be associated with phase transition like 
phenomenon
and self organization in the plasma sheet \cite{chan99, cons01, conl01, milo01}.
Our results also support the idea of \\
\cite{ang99a} that 
BBF--related intermittent turbulence may represent an effective way for CSC. Propagating BBFs can modify 
a critical threshold for nonlinear instabilities or trigger further localized reconnections because of
the free energy present on multiple  scales. In this sense, our results suggest that BBFs may represent
those multiscale carriers of energy, flux and momentum, which lead to the avalanche-like spread of 
disturbances on medium or large-scales \cite{klim00, lui02}. In this respect classification 
of multi-scale physical processes using $LIM$, or multifractal distributions 
offers a way in which the role of turbulence 
in a variety of dynamical processes within plasma sheet can be statistically evaluated.
\\
Acknowledgements\\
The authors acknowledge the use of FRACLAB package developed at the Institut National de 
Recherche en Informatique, France. ZV thanks A. Petrukovich for many valuable suggestions.
The work by KHG was financially supported by the German Bundesministerium f\"{u}r Bildung 
und Wissenschaft and the German Zentrum  f\"{u}r Luft- und Raumnfahrt under contract 50 OC 0103.
\\


\pagebreak
\begin{figure}[tb]
\centerline{
\includegraphics[width=3.2in]{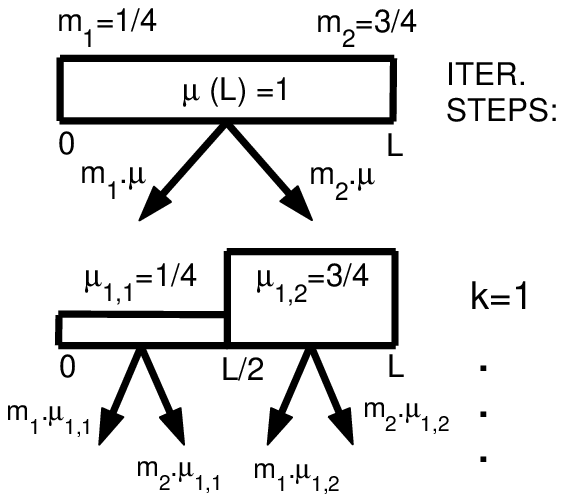}
}
\caption{Recursive construction rule for binomial distribution}
\end{figure}

\begin{figure}[tb]
\centerline{
\includegraphics[width=3.2in]{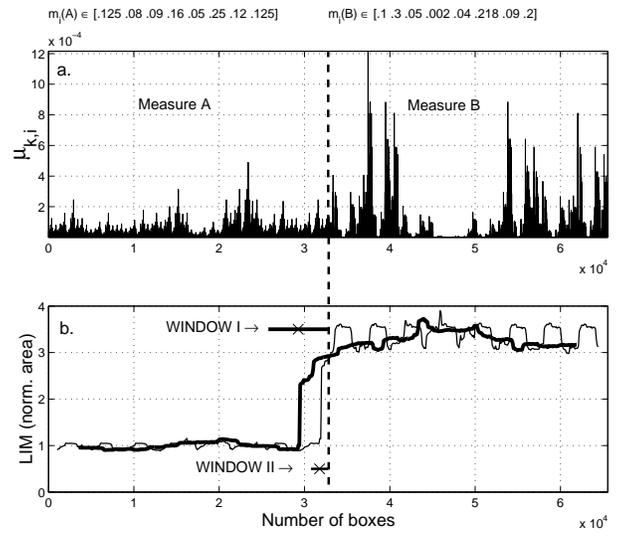}
}
\caption{ a. Two multinomial distributions: measure A is less intermittent than
 measure B; dashed line in the middle
separates the two measures. b. $LIM$ estimation for two different windows I and II.}
\end{figure}

\begin{figure}[tb]
\centerline{
\includegraphics[width=3.2in]{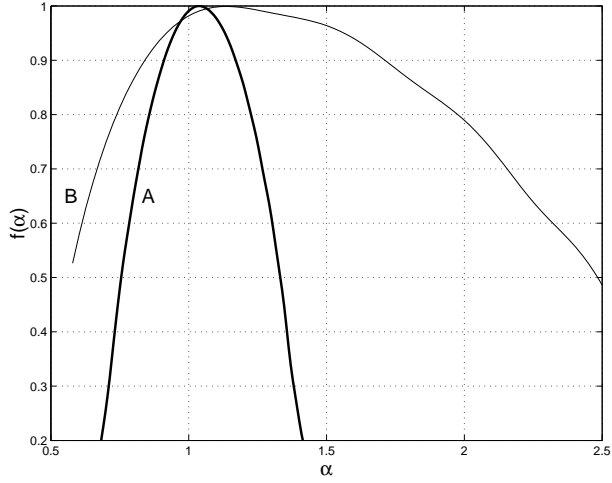}
}
\caption{ Multifractal distributions for measures A and B shown in Figure 2a.}
\end{figure}

\begin{figure}[tb]
\centerline{
\includegraphics[width=3.2in]{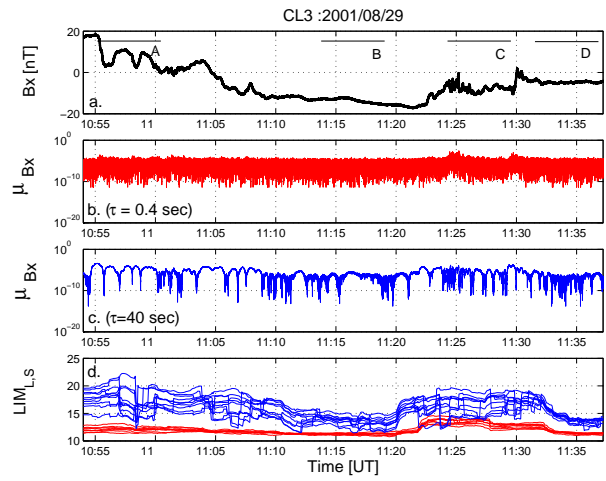}
}
\caption{a. Magnetic field $B_X$ component measured by Cluster 3; b. the associated measure computed by using 
Equations 2,3  on small scales (red colour); c. the same on large scales (blue colour); d. small and large
scale $LIM_{L,S}$.}
\end{figure}

\begin{figure}[tb]
\centerline{
\includegraphics[width=3.2in]{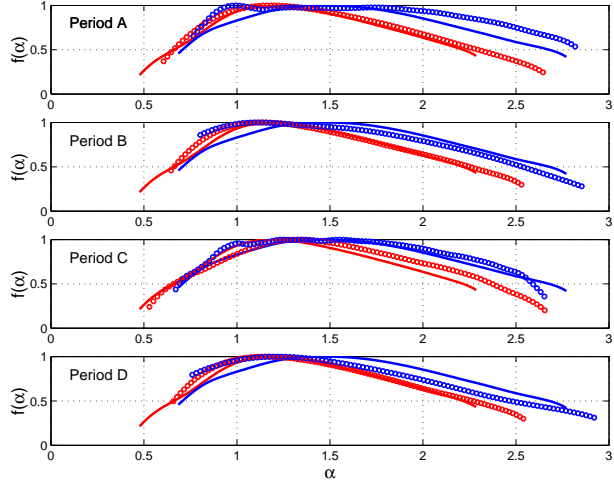}
}
\caption{Multifractal spectra for periods A--D shown in Figure 4a (red circles: small scales, blue circles: large
scales); continuous curves with the same colour code correspond to average multifractal spectra estimated
for the whole interval from 1055 to 1135 UT.}
\end{figure}

\begin{figure}[tb]
\centerline{
\includegraphics[width=3.2in]{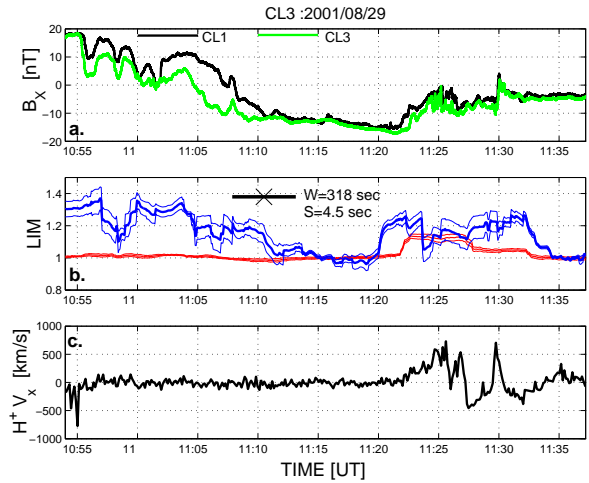}
}
\caption{a. Magnetic field $B_X$ components measured by Cluster 1, 3 spacecraft; b.$LIM_{L,S}$ for small
scales (red line) and large scales (blue line), thin curves show standard deviations; 
c. proton velocity data.}
\end{figure}

\begin{figure}[tb]
\centerline{
\includegraphics[width=3.2in]{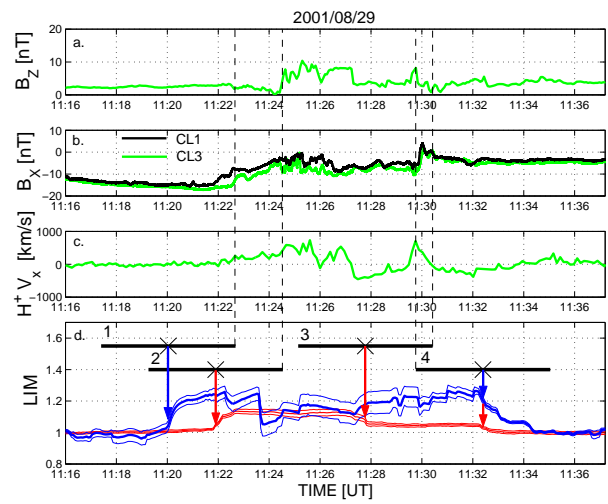}
}
\caption{a. Magnetic field $B_Z$ component from Cluster 3; b. Magnetic field $B_X$ components from Cluster 1, 3; 
c. proton velocity data; d. $LIM_{L,S}$}
\end{figure}


\end{document}